\documentclass[aps,twocolumn,amsmath,amssymb, superscriptaddress]{revtex4}
\usepackage[pdftex]{graphicx}% Include figure files
 \usepackage{amsmath}
\usepackage[T1]{fontenc}
\usepackage{amssymb}
\usepackage{amsfonts}
\usepackage{bm}
 \usepackage{amsmath} 
\usepackage{lipsum}
\usepackage{amsfonts} 
\usepackage{amssymb, mathrsfs}
\usepackage{braket}
\usepackage{graphicx} 
\usepackage{subfigure}
\usepackage{bbm}
\def\beq{\begin{equation}}
\def\eeq{\end{equation}}
\def\bsp{\begin{split}}
\def\esp{\end{split}}
\def\bea{\begin{eqnarray}}
\def\eea{\end{eqnarray}}
\def\ba{\begin{array}}
\def\ea{\end{array}}

\def\dg{\dagger}
\def\sg{\sigma}

\def\lb{\left(}
\def\rb{\right)}

\def\l.{\left.}
\def\r.{\right.}

\def\ra{\rangle}
\def\la{\langle}

\def\bo{\bold{k}}

\begin{document}

%\date{\today}
\title{Magnon edge states in hardcore-Bose-Hubbard model}
\author{S. A. Owerre}
\affiliation{Perimeter Institute for Theoretical Physics, 31 Caroline St. N., Waterloo, Ontario N2L 2Y5, Canada.}
\affiliation{African Institute for Mathematical Sciences, 6 Melrose Road, Muizenberg, Cape Town 7945, South Africa.}
\email{solomon@aims.ac.za}

\begin{abstract}
Quantum Monte Carlo  (QMC)  simulation has uncovered nonzero Berry curvature and bosonic edge states in hardcore-Bose-Hubbard model on the gapped honeycomb lattice. The competition between the chemical potential and  staggered onsite potential leads to  an interesting quantum phase diagram comprising  superfluid phase, Mott insulator, and charge density wave insulator.     In this paper, we present a semiclassical perspective of this system by mapping to a spin-$1/2$ quantum XY model.  We give an explicit analytical origin of the quantum phase diagram, the Berry curvatures,  and the edge states using  semiclassical approximations.    We find very good agreements between the semiclassical analyses and the QMC results. Our results show that   the  topological properties of hardcore-Bose-Hubbard  model are the same as those of magnon in the  corresponding quantum spin system. Our results are applicable to systems of ultracold bosonic atoms  trapped in honeycomb optical lattices. \end{abstract}
\maketitle

 \section{Introduction}
We are used to  the topological properties of fermion band theory in electronic systems, which have been studied extensively over the past decade \cite{sem1, sem, yu, yu1, yu2, yu3, yu4, yu5, yu6, yu7, yu8, fdm}.  Recently, the study of topological band theory has been extended to bosonic systems. A natural extension of topological properties of fermionic systems to bosonic systems can be achieved by replacing the lattice sites of fermions with bosons \cite{pa1,var1}. Hence, the fermionic operators can be regarded as bosonic operators obeying a different commutation relation.  

In the hardcore limit, the bosonic systems  map to  spin-$1/2$ quantum magnets  \cite{matq}.  This correspondence is very crucial as it paves the way to interpret results in terms of bosons as well as spin variables. It also means that the excitations of hard-core bosons must be the underlying spin wave excitations (magnons) of the corresponding quantum spin model. Therefore, the topological properties of the  bosonic excitations must be similar to that of spin wave excitations.  In this regard, the Haldane  spin-orbit coupling \cite{fdm} in the hardcore limit maps to an out-of-plane Dzyaloshinskii-Moriya interaction (DMI), which induces a nonzero Berry curvature and thermal Hall effect of magnetic spin excitations \cite{alex1, alex0, alex2,alex5,alex4, sol1,sol, alex5a, alex6,  alex1a}.   However, in contrast to fermionic systems, there is neither a  Fermi energy nor a filled band in bosonic systems. This means that the topological invariant quantity usually called the Chern number must be independent of the statistical nature of the particles. It simply predicts the existence of edge state modes in the vicinity of the bulk energy gap as a result of the bulk-edge correspondence. This leads to edge states in bosonic systems.

Unfortunately, many topological bosonic models have a numerical sign problem that hinders an explicit quantum Monte Carlo (QMC) simulation due to an imaginary statistical average. In a recent study,  Guo {\it et al}~ \cite{alex9} have investigated the Bose-Hubbard model on the honeycomb lattice using QMC. This model is devoid of the debilitating QMC sign problem as there is no imaginary phase amplitude. It is analogous to fermionic graphene model without spin-orbit coupling in the presence of a biased potential \cite{sem,sem1}.   The authors  have explicitly mapped out the bosonic quantum phase diagram, the Berry curvature and edge states characterizing the topological properties of the system induced by a staggered on-site potential.

In this paper, we present another perspective of their QMC results using a semiclassical approach.    The QMC results presented in Ref.~\cite{alex9} utilized the electronic analogue of the Bose-Hubbard  model. Here, we show that the entire QMC analysis can be understood semi-classically.  This is due to the fact that the  hardcore-Bose-Hubbard  model is merely a spin-$1/2$ quantum XY model with  competing sublattice magnetic fields,  thus the results can also be interpreted in terms of magnetic spins and the semiclassical approach is known to be suitable for such models \cite{ber, tom}.   We find that the quantum phase diagram uncovered by QMC can actually  be understood by mean-field theory. We uncover the same three insulating phases:  superfluid (SF), Mott insulator phase, and charge-density-wave (CDW) insulator.  The latter insulating phase is a consequence of the competing sublattice magnetic fields. 

As mentioned above, the correspondence between hard-core bosons and quantum spin systems suggests that the spin wave excitations correspond to the bosonic excitations.     As QMC showed, the topological properties of this system is manifested by a nonzero Berry curvature. We  show that the Berry curvature  of the magnon excitations in the $\rho=1/2$ CDW insulator has the same trend as the one obtain by QMC simulation \cite{alex9}. In contrast to DMI induced edge states with nonzero Chern number \cite{alex1, alex0, alex2,alex5,alex4, sol1,sol, alex9, alex5a, alex6,  alex1a}, the Chern number of the present model vanishes. Nevertheless,  we observe zigzag bosonic magnon edge states which do not have the same origin as those in DMI system \cite{sol}. It is noted that nontrivial topology has been realized in two-dimensional (2D) optical fermionic \cite{jot} and bosonic \cite{jot1, jot2} atoms. Thus, our results are applicable to these systems.

 \section{Hardcore-Bose-Hubbard model}
\begin{figure}[ht]
\centering
\includegraphics[width=1.75in]{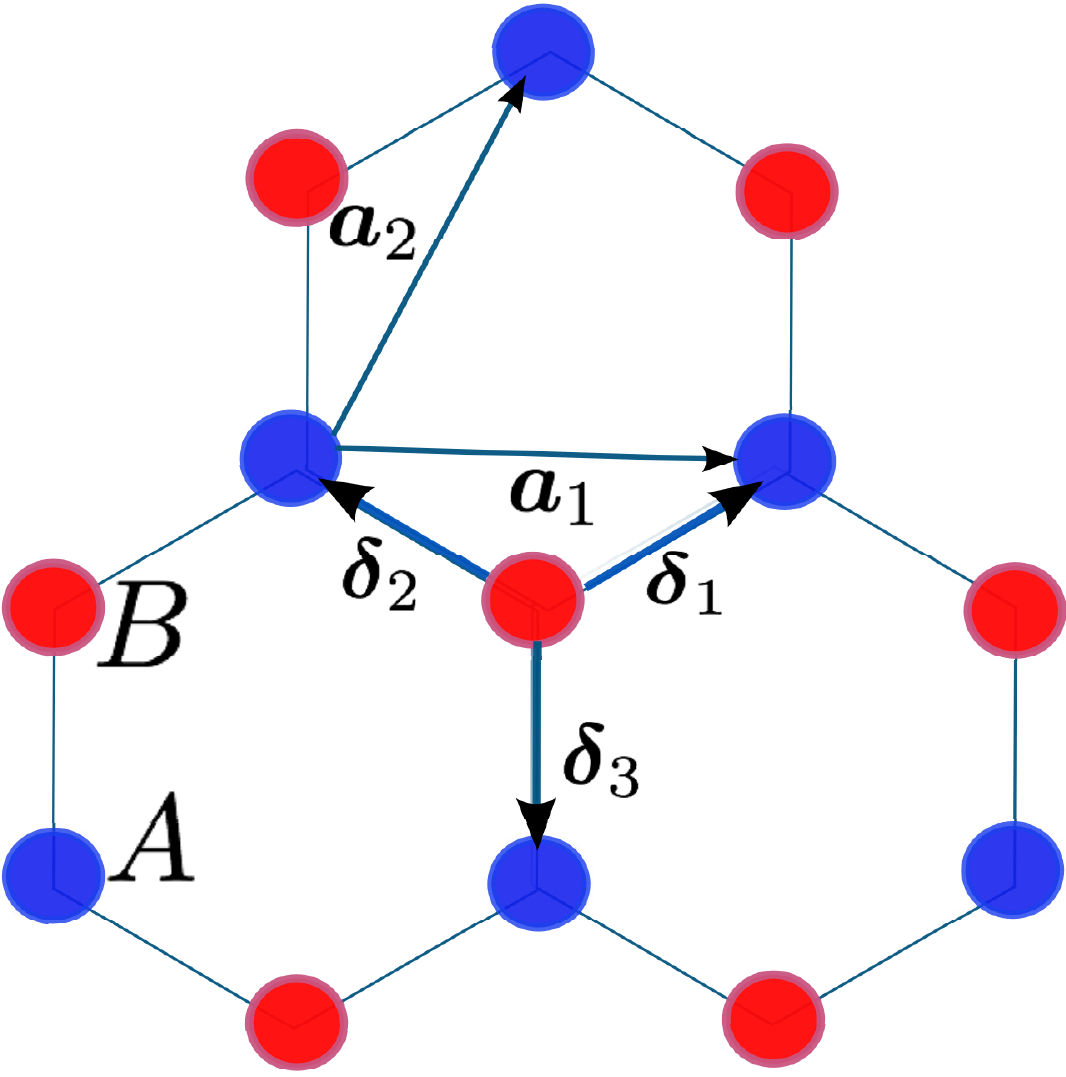}
\caption{Color online. The  honeycomb lattice with two sublattices $A$ and $B$ indicated by different  colors.  The coordinates are $\bold a_1=\sqrt{3}a\hat x;~ \bold a_{2}=a(\sqrt{3}\hat x, 3\hat y)/2$; $ \boldsymbol{\delta}_{1,2}=a(\pm\sqrt{3}\hat x,~\hat y)/2$, and $ \boldsymbol{\delta}_3=a(0, -\hat y)$. }
\label{unit}
\end{figure}

 A recent QMC simulation by Guo {\it et al} ~\cite{alex9} studied the topological properties of the extended harcdore-Bose-Hubbard model governed  by the Hamiltonian
 \begin{align}
H&= -t\sum_{\langle ij\rangle}( b^\dg_i b_j +  h.c.) +\sum_i U_i n_i -\mu\sum_i  n_i\label{hardcore},
\end{align}
where,  $t>0$ denotes NN hopping, $\mu$ is the chemical potential, and $U_i$ is a staggered on-site potential , with $U_i=\Delta$ on sublattice $A$, and $U_i=-\Delta$ on sublattice $B$ of the honeycomb lattice shown in Fig.~\ref{unit}.  $n_i=b^\dg_ib_i$,  $b^\dg_i$ and $ b_i$ are the bosonic creation and annihilation operators respectively. They obey the algebra $[b_i, b_j^\dg]=0$ for $i\neq j$ and $\lbrace b_i, b_i^\dg \rbrace=1$. 
\begin{figure}[ht]
\centering
\includegraphics[width=3.5in]{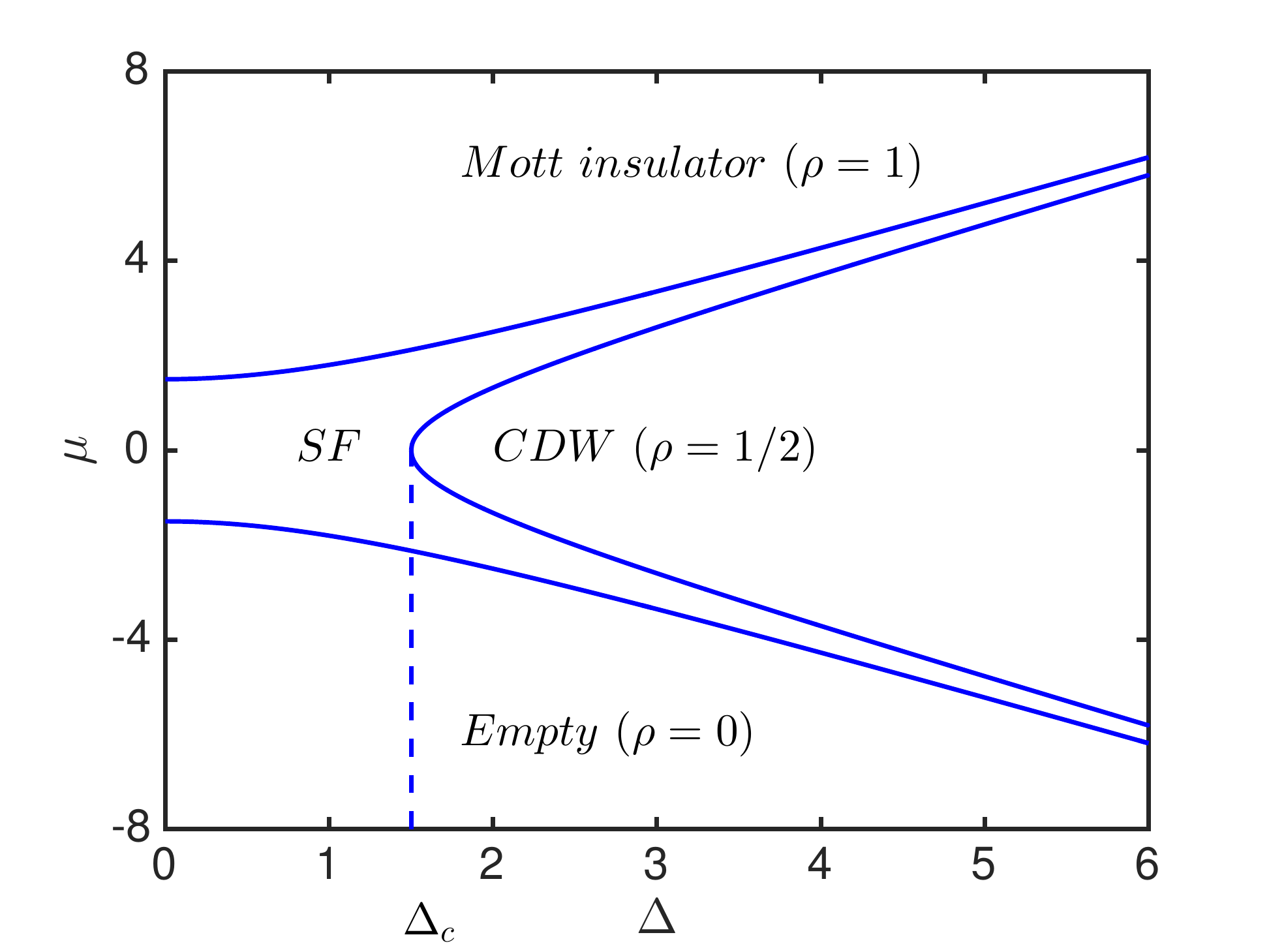}
\caption{Color online. Mean-field phase diagram of the Bose-Hubbard model \ref{hardcore}, where $J=1/2$ is the parameter value in the spin language, which corresponds to $t=1$ in the hard-core bosons. The dash line indicates the critical value of the $CDW$ phase.}
\label{phase}
\end{figure}

%\begin{figure}
%\centering
%\includegraphics[width=3.5in]{band_CDW.png}
%\caption{Color online. Berry curvatures and the magnon bulk bands of the Bose-Hubbard model at $J=1/2$,  $\mu=2J$, $\Delta=4J$. This corresponds to the $\rho=1/2$ CDW insulating phase in Fig.~\ref{phase}. The equal and opposite maximum peaks of the Berry curvatures are consistent with  QMC simulation \cite{alex9}.  }
%\label{band_berry}
%\end{figure}
%\begin{figure}
%\centering
%\includegraphics[width=3.5in]{band_SF.png}
%\caption{Color online. Berry curvatures and the magnon bulk bands of the Bose-Hubbard model at $J=1/2$,  $\mu=J$, $\Delta=J$. This corresponds to the superfluid (SF) phase in Fig.~\ref{phase}. The Berry curvatures in this phase are not measured in QMC simulation \cite{alex9}.  }
%\label{band_sf}
%\end{figure}

For fermionic systems,  the momentum space Hamiltonian for $\mu=0$ is given by
\begin{align}
\mathcal{H}(\bold k)=\left(
\begin{array}{cc}
\Delta&-tf_\bo \\
-tf^*(\bo)&-\Delta
\end{array}
\right),
\label{honn}
\end{align}
where $f_\bo=e^{ik_ya/2}\lb 2\cos(\sqrt{3}k_xa/2)+e^{-3ik_ya/2}\rb$.
 %Also the structure factor can be written as $f_\bo=1+e^{i\bo\cdot{\bf a}_1}+e^{i\bo\cdot{\bf a}_2}$
 The corresponding eigenvalues are
\begin{align}
\epsilon_\pm(\bo)=\pm\sqrt{\Delta^2+t^2|f_\bo|^2}.
\label{fen}
\end{align}
In the bosonic version, the energy does not have this simple symmetric form as we will show later. In this case, we adopt the quantum spin analogue of the Bose-Hubbard Hamiltonian  \ref{hardcore}  via the Matsubara-Matsuda transformation \cite{matq},  $S_i^+ \to b^\dagger_i,~S_i^-\to b_i,~S_i^z\to n_i-1/2$. The resulting quantum spin Hamiltonian is given by
 \begin{align}
H&=-J\sum_{\langle ij\rangle}(S_i^+S_j^-+ S_i^-S_j^+)-\sum_{i} (\mu- U_i) S_i^z,
\label{hh1}
\end{align}
where $S_i^{\pm}=S_i^x\pm iS_i^y.$ The last term is basically a competing magnetic field on the two sublattices.   Throughout the analysis in this paper we fix $J=1/2$, which corresponds to $t=1$ in the hard-core bosons.
\section{Mean-field phase diagram}
In this section, we present  the mean-field phase diagram of the Bose-Hubbard model \ref{hardcore}. The mean-field approximation is implemented by approximating the spins  as classical vectors parameterized by a unit vector: $\bold{S}_i=S\lb\sin\theta_i\cos\phi_i, \sin\theta_i\sin\phi_i,\cos\theta_i \rb$. We adopt the customary  two-sublattice  honeycomb lattice depicted in Fig.~\ref{unit}. Since the spins lie on the same plane we take  $\phi_i=0$, then the classical energy  is parameterized by $\theta_i$  given by
\begin{align}
e_c &= -\Delta_c \sin\theta_A\sin\theta_{B}-(\mu-\Delta)\cos\theta_A -(\mu+\Delta)\cos\theta_B, 
\label{cla}
\end{align}
where $e_c=E_c/NS$, $\Delta_c=2JzS$, $S=1/2$, $N$ is the number of unit cells, and $z=3$ is the coordination number of the lattice. The filling factor is given by $\rho= 1/2 + S(\cos\theta_A +\cos\theta_B)/2$. 
%\begin{figure}[ht]
%\centering
%\includegraphics[width=3.5in]{angle}
%\caption{Color online. The classical trajectory of the hardcore Bose-Hubbard model.}
%\label{angle}
%\end{figure}
 There are three phases in this model uncovered by QMC \cite{alex9}.  In the mean-field approximation they are characterized by the polar angles. In the SF phase $\theta_A=\theta_B\neq 0,\pi$, the Mott phase is characterized by $\theta_A=\theta_B=0$ or $\pi$, and the CDW is characterized by $\theta_A=0;~\theta_B=\pi$ or $\theta_A=\pi;~\theta_B=0$. The mean field phase diagram is derived by minimizing \ref{cla} following the standard approach \cite{kle}.  We obtain the phase boundary between SF and Mott insulators as $\mu_{c1}=\pm\sqrt{\Delta^2+\Delta^2_c}$. This corresponds to $\bo=0$ in Eq.~\ref{fen}. The phase boundary between SF and CDW insulators  is $\mu_{c2}=\pm\sqrt{\Delta^2-\Delta^2_c}$. Unlike the first phase boundary, this expression cannot be obtained from Eq.~\ref{fen}. The classical angles are obtained explicitly as
\begin{align}
\cos^2\theta_A&=\lb\frac{\Delta-\mu}{\Delta_c}\rb^2\bigg[\frac{(\Delta+\mu)^2+\Delta_c^2}{(\Delta-\mu)^2+\Delta_c^2}\bigg]\label{eq1},\\
\cos^2\theta_B&=\lb\frac{\Delta+\mu}{\Delta_c}\rb^2\bigg[\frac{(\Delta-\mu)^2+\Delta_c^2}{(\Delta+\mu)^2+\Delta_c^2}\bigg].
\label{eq2}
\end{align}
The  mean-field phase diagram is depicted in Fig.~\ref{phase}.  The superfluid phase appears  for small $\mu$, whereas the Mott phase is predominant for large $\mu$. The CDW arises mainly from the competition between $\mu$ and $\Delta$. The threshold limit  $\Delta_c$ corresponds to the point where $\mu_{c2}=0$.  This is the exact same quantum phase diagram uncovered by QMC \cite{alex9}.

\section{Band structure}
The main purpose of this paper is to show that the magnon excitation of the  Bose-Hubbard model \ref{hardcore} embodies the topological properties of this system. Since the Bose-Hubbard model \ref{hardcore} describes an ordered system as shown in the phase diagram Fig.~\ref{phase},  we can study the excitations of the spin waves when quantum fluctuations are introduced and this should correspond to the excitations of the bosons as explained above. The simplest way to study spin wave excitations is via the standard Holstein Primakoff transformation. This approach is frequently used in the study of hard-core bosons in two-dimensional lattices \cite{ber,tom}.  In term of topological properties of quantum magnets, the Holstein-Primakoff transformation has also been utilized effectively in this regard \cite{alex1, alex0, alex2,alex5,alex4, sol1},  and considered to be a good experimental predictor \cite{alex6}.  In this section,  we utilize this semiclassical formalism in the study of the Bose-Hubbard model. The starting point of spin wave expansion is the rotation of the coordinate axes such that the $z$-axis coincides with the local direction of the classical polarization.  This is implemented by a rotation about the $y$-axis on the two sublattices
\begin{figure}[!]
\centering
\includegraphics[width=1\linewidth]{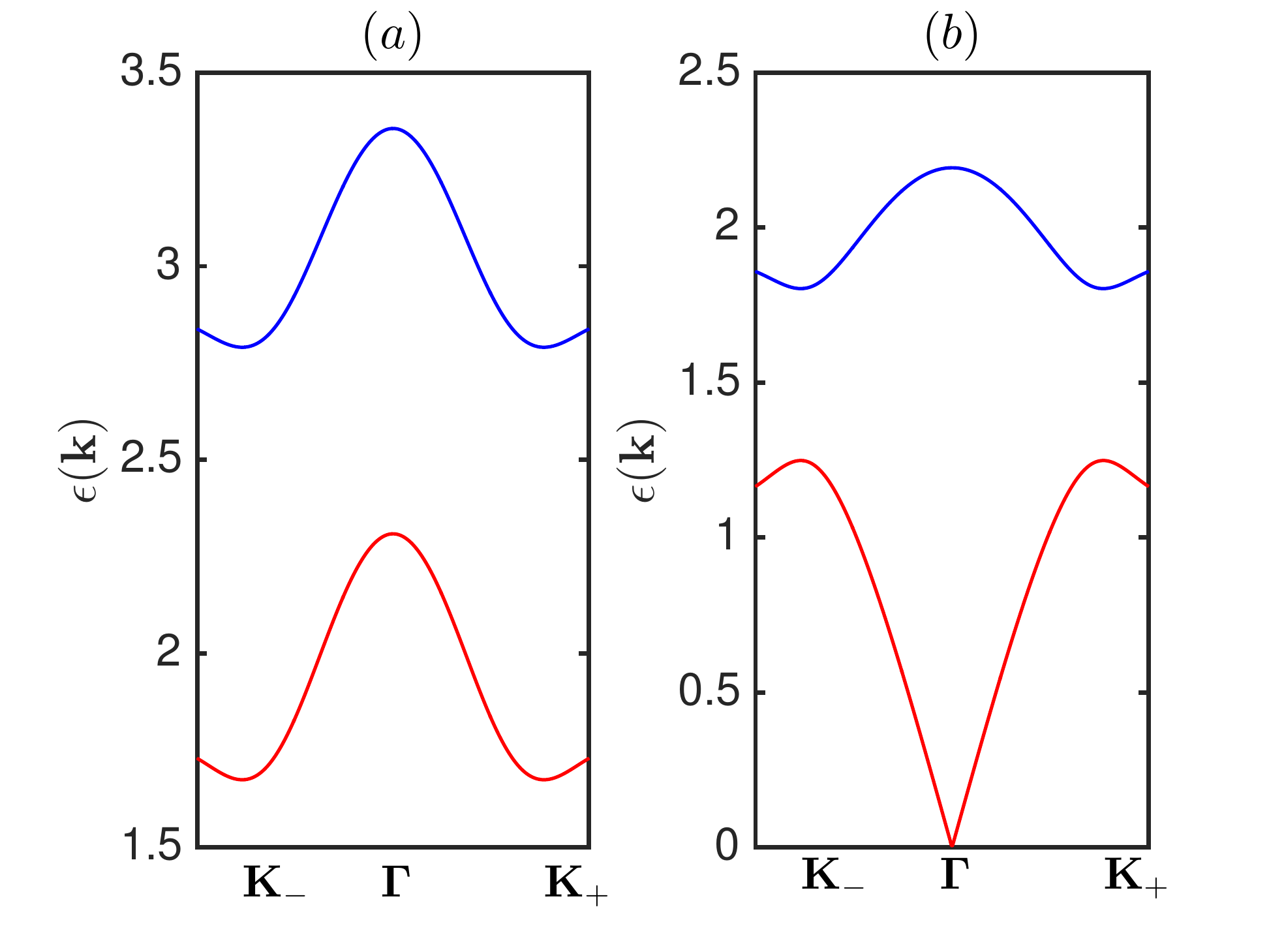}
\caption{Color online.  Bosonic magnon bands of the Bose-Hubbard model.  $(a)$ CDW phase $J=1/2$,  $\mu=2J$, $\Delta=4J$.  $(b)$ SF phase $J=1/2$,  $\mu=J$, $\Delta=J$. }
\label{band}
\end{figure}

\begin{align}
&S_{i\alpha}^x=S_{i\alpha}^{\prime x}\cos\theta_\alpha  +  S_{i\alpha}^{\prime z}\sin\theta_\alpha,\label{trans}\nonumber\\&
S_{i\alpha}^y=S_{i\alpha}^{\prime y},\\&\nonumber
S_{i\alpha}^z=- S_{i\alpha}^{\prime x}\sin\theta_\alpha + S_{i\alpha}^{\prime z}\cos\theta_\alpha,
\end{align}
where $\alpha=A,B$ label the sublattices.

We then introduce  the linearized Holstein Primakoff transformation, $S_{i\alpha}^{\prime z}= S-c_{i\alpha}^\dagger c_{i\alpha},~
 S_{i\alpha}^{\prime y}=  i\sqrt{ S/2}(c_{i\alpha}^\dagger -c_{i\alpha}),~
 S_{i\alpha}^{\prime x}=  \sqrt{S/2}(c_{i\alpha}^\dagger +c_{i\alpha})$. The bosonic tight binding Hamiltonian becomes
 \begin{align}
H&=-\sum_{\la ij\ra}[v_{1}( c_{iA}^\dagger c_{jB}+ h.c.) +v_{2}( c_{iA}^\dagger c_{jB}^\dagger+ h.c.)] \nonumber\\& +(v_A- m_A)\sum_i c_{iA}^\dagger c_{iA}+(v_B+m_B)\sum_j c_{jB}^\dagger c_{jB},
\label{hp3}
\end{align}
where $v_{1,2}=JS(\cos\theta_A\cos\theta_B \pm 1)$, $v_{A/B}=\Delta_c\sin\theta_A\sin\theta_B+\mu\cos\theta_{A/B}$ and $ m_{A/B}=\Delta\cos\theta_{A/B}$. Apart from the off-diagonal terms with coefficient $v_2$, Eq.~\ref{hp3} is similar to a graphene model with  a staggered potential.   The energy bands are given in Appendix \ref{appen1}.  Figure~\ref{band}(a) shows the magnon bands  in the $\rho=1/2$ CDW insulator and  Fig.~\ref{band}(b) shows the magnon bands  in the SF phase. We see that the lower band in the SF phase has a Goldstone model at $\bo =0$ (see Appendix \ref{appen3}) in contrast to the CDW insulator. A special limit of the CDW insulator is analyzed in Appendix \ref{appen2}. It is noted that there are two SF phases in this model--- gap SF phase for $\Delta\neq 0$ and gapless SF phase for $\Delta=0$ (see Appendix \ref{appen3}).

\section{Magnon edge states}
 To study the topological properties of this model we use the results in Appendix \ref{appen1}.    For $\Delta=0$, we have $m_A=m_B=0$, then Eqs.~\ref{eq1} and \ref{eq2} simply give $\theta_A=\theta_B=\theta$, hence $v_A=v_B$ and  the system reduces to the usual hard-core bosons or XY model. In this limit,  the system exhibits Dirac nodes at  $\bold K_\pm=(\pm 4\pi/3\sqrt{3}a, 0)$ and a Goldstone mode at ${\bf \Gamma}=0$.  As QMC demonstrated \cite{alex9},  the   topological properties of this system is induced by a nonzero $\Delta$ which plays the role of a gap as shown in Appendix \ref{appen1}. This  implies that $m_A\neq 0$ and $ m_B\neq 0$.
\begin{figure}
\centering
\includegraphics[width=3.5in]{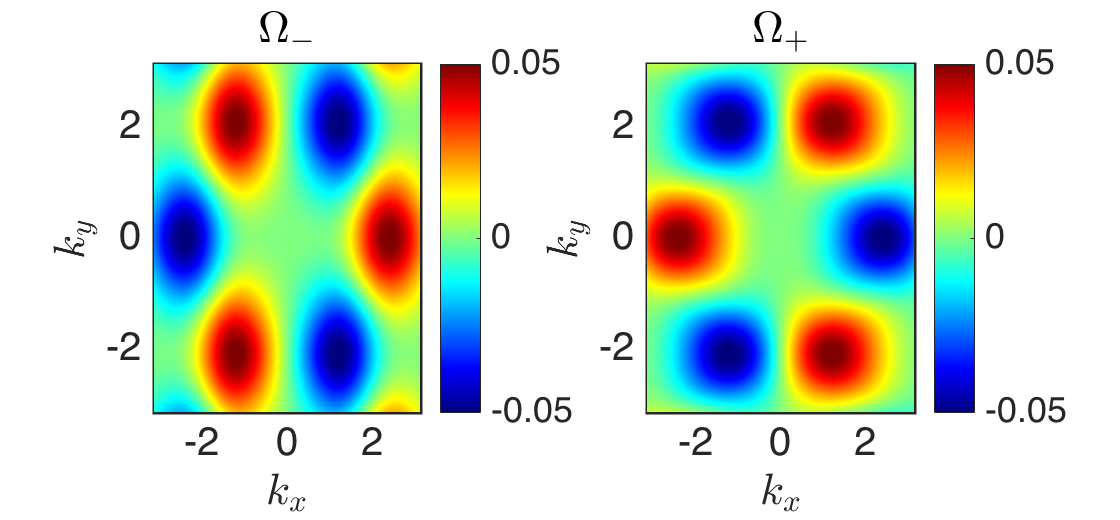}
\caption{Color online. Berry curvatures  of the Bose-Hubbard model at $J=1/2$,  $\mu=2J$, $\Delta=4J$. This corresponds to the $\rho=1/2$ CDW insulating phase in Fig.~\ref{phase}. The minima and maxima of the Berry curvatures are consistent with  QMC simulation \cite{alex9}.  }
\label{bc_berry}
\end{figure}
\begin{figure}
\centering
\includegraphics[width=3.5in]{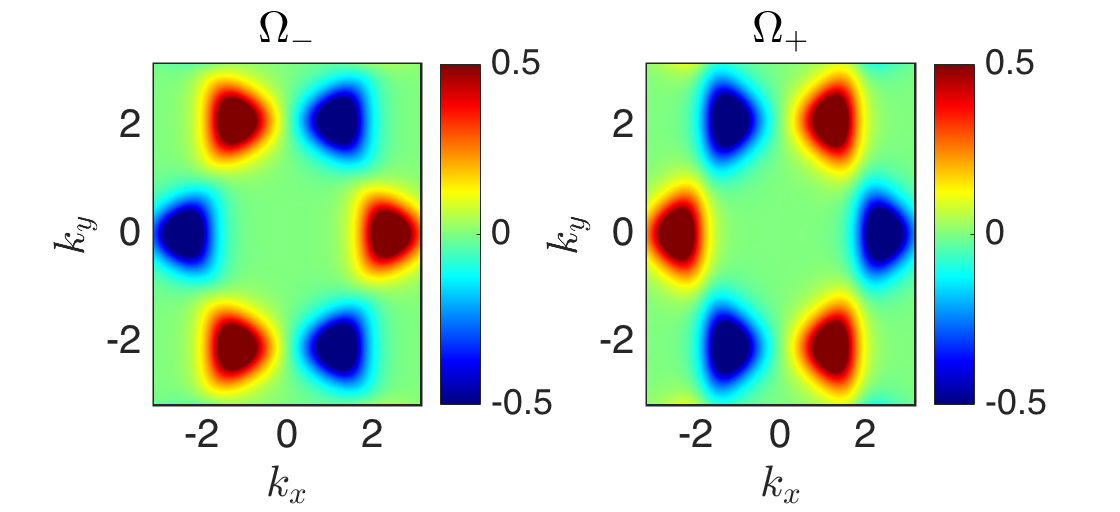}
\caption{Color online. Berry curvatures  of the Bose-Hubbard model at $J=1/2$,  $\mu=J$, $\Delta=J$. This corresponds to the superfluid (SF) phase in Fig.~\ref{phase}. The Berry curvatures in this phase are not measured in QMC simulation \cite{alex9}.  }
\label{bc_sf}
\end{figure}
\begin{figure}
\centering
\includegraphics[width=3.5in]{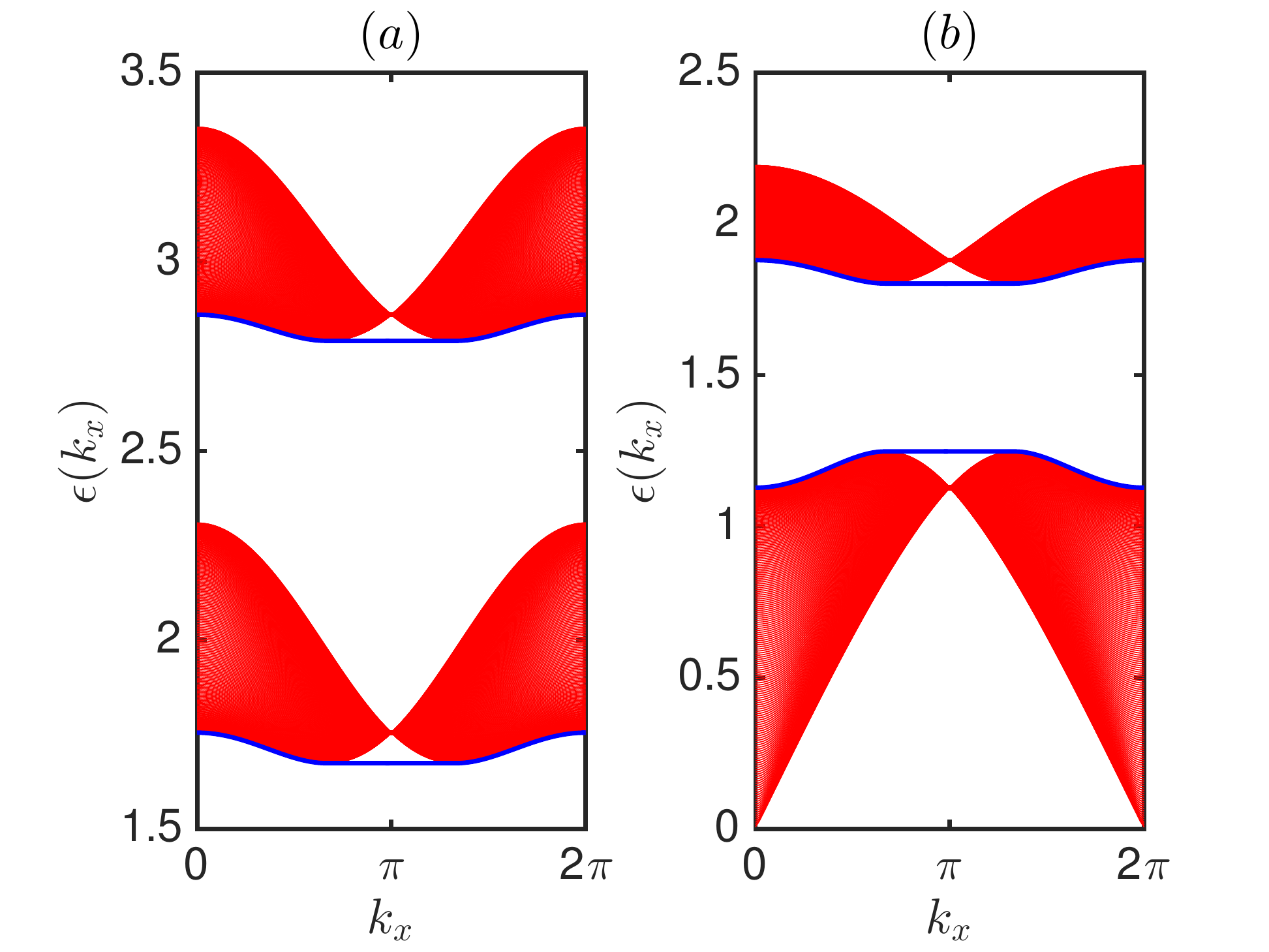}
\caption{Color online. Magnon zigzag edge states (green solid lines)  of the Bose-Hubbard model in CDW $(a)$ and SF $(b)$ phases. The parameters are the same as Fig.~\ref{band}. The structure of the bands and the edge states are consistent with QMC simulation \cite{alex9}.  }
\label{edge}
\end{figure}

We are interested in the Berry curvature associated with the magnon bulk gap. It is given by
\begin{align}
\Omega_\lambda(\bold k)=-\sum_{\lambda\neq \lambda^\prime}\frac{2\text{Im}[ \braket{\mathcal{U}_{\bo\lambda}|v_x|\mathcal{U}_{\bo\lambda^\prime}}\braket{\mathcal{U}_{\bo\lambda^\prime}|v_y|\mathcal{U}_{\bo\lambda}}]}{\lb\epsilon_{\bo\lambda}-\epsilon_{\bo\lambda^\prime}\rb^2},
\label{chern2}
\end{align}
where   $v_{i}=\partial \mathcal{H}_B(\bold k)/\partial k_{i}$ defines the velocity operators, $\mathcal{U}_{\bo\lambda}$ denotes the columns of the matrix that diagonalizes $\mathcal{H}_B(\bold k)$(see Appendix~\ref{appen1}), and  $\lambda=\pm$ denotes the two positive magnon bands.  The CDW and the SF phases are the nontrivial phase in this model. Figure~\ref{bc_berry} shows the Berry curvatures for the top and the bottom bands in the $\rho=1/2$ CDW insulator and Fig.~\ref{bc_sf} shows the Berry curvatures in the SF phase.  The Berry curvatures show minima and maxima peaks at the corners of the Brillouin zone (see Appendix~\ref{appen2}). This is in good agreement with QMC simulation \cite{alex9}. In contrast to DMI induced Berry curvatures, the Chern number of each band $\mathcal{C}_\lambda= \frac{1}{2\pi}\int_{{BZ}} d^2k~ \Omega_\lambda(\bold k),$ vanishes identically for the present model \cite{footnote}. However, due to  nonzero Berry curvatures  we observe zigzag edge states in this system  for $k_x\in[2\pi/3\sqrt{3}, 4\pi/3\sqrt{3}]$ as depicted in Fig.~\ref{edge}. Thus, they have a different origin from the DMI induced ones \cite{sol}.  This is consistent with QMC simulation of Ref.~\cite{alex9}.  
% Another important observation is that the Berry curvature  The existence of edge  states in these insulators is already encoded in the non-vanishing Berry curvatures.  They can be determined in the usual way as we have previously shown with a model in which QMC has a sign problem \cite{sol2}.  Although, the mass term in the present model is momentum independent, the idea is essentially the same. From Eq.~\ref{hp3}, it is evident that the magnon tight binding model is analogous to that of graphene, so one could follow a similar procedure as demonstrated in  Ref.~\cite{sem,sem1} to solve for the edge states and a similar situation still emerge.   

\section{Conclusion}

We have complemented the QMC simulation of Guo {\it et al}~ \cite{alex9} using a semiclassical approach.  The main result of our study is that the topological properties of hard-core bosons correspond to the topological properties of the magnon bulk bands of the corresponding quantum spin model.  In the hardcore-Bose-Hubbard model, competing sublattice magnetic fields lead to a nontrivial charge-density-wave insulator with a filling factor of $\rho=1/2$, in addition to superfluid phase and Mott insulator. We have uncovered the mean-field phase diagram,  which is consistent with the QMC phase diagram. We also derived the magnon energy bands of each phase and show that the corresponding Berry curvatures and edge states are consistent with QMC simulations.   This basic idea we have presented here can also be generalized to bilayer honeycomb lattice. These results will be useful in experimental set up of ultracold bosonic atoms in honeycomb optical lattice.

\section*{Acknowledgments}
The author would like to thank African Institute for Mathematical Sciences (AIMS). Research at Perimeter Institute is supported by the Government of Canada through Industry Canada and by the Province of Ontario through the Ministry of Research
and Innovation.
  
\appendix
%\begin{figure*}[!]
%\centering
%  \subfigure[\label{nb}]{\includegraphics[width=.45\linewidth]{Neel_band}}
%   \quad
%   \subfigure[\label{nb1}]{\includegraphics[width=.4\linewidth]{Neel_edge}}
%\caption{Color online.  Bosonic magnon band  $(a)$  and edge state (green solid line) $(b)$ at $\mu=0,~\Delta=\Delta_c$, corresponding to fully polarized N\'eel state.}
%\label{Nband}
%\end{figure*}
\section{General diagonalization}
\label{appen1}
In this appendix, we show step by step diagonalization of the Hamiltonian in the text above. 
In momentum space,  we introduce the Nambu operators $\Psi^\dg_\bo= (\psi^\dg_\bo, \thinspace \psi_{-\bo} )$, with  $\psi^\dg_\bo=(c_{\bo A}^{\dg},~\thinspace c_{\bo B}^{\dg})$ and write Eq.~\ref{hp3} as $ H= \frac{1}{2}\sum_{\bo}\Psi^\dg_\bo \cdot \mathcal{H}(\bo)\cdot\Psi_\bo + \text{const.},$ 
where
\begin{align}
\mathcal{H}(\bo)=
\begin{pmatrix}
v_A-m_A&-v_1f_\bo& 0&-v_2f_\bo\\
-v_1f^*_\bo& v_B+m_B&-v_2f^*_\bo&0\\
0&-v_2f_\bo& v_A-m_A&-v_1f_\bo\\
-v_2f_\bo^*&0&-v_1f_\bo^*&v_B+m_B
\end{pmatrix},
\label{mat}
\end{align}
 which can be written in compact form as
 \begin{align}
\mathcal{H}(\bo)&= [{ (v_A- m_A)\tau_+^0} +(v_B+m_B){\tau_-^0}]\otimes{\sigma}_0 \nonumber\\&- v_{1}(\tau_+f_\bo +h.c.)\otimes{\sigma}_0-  v_{2}(\tau_+f_\bo +h.c.)\otimes\sigma_x.
\label{hhhm}
\end{align}
 We have introduced two Pauli matrices  $\boldsymbol\sigma$ and $\boldsymbol\tau$, where $\tau_\pm= (\tau_x\pm i\tau_y)/2$, while  ${\sg}_0$ is an identity $2\times 2$ matrix in the $\boldsymbol\sigma$-space. We also introduce other matrices for simplification, $\tau^0_+=\text{diag}(1,0)$, $\tau^0_-=\text{diag}(0,1)$. 

 The spin wave Hamiltonian  is Hermitian but it is not diagonal. To diagonalize this Hamiltonian, we make a transformation $\Psi^\dg_\bo\to\mathcal{U}_\bo{\Psi}^\dagger_\bo$, which satisfies the relation \bea\mathcal{U}_\bo^\dg \mathcal{H}(\bo) \mathcal{U}_\bo= \epsilon(\bo); \quad \mathcal{U}_\bo^\dg \eta \mathcal{U}_\bo= \eta,\eea  with $\eta=\sigma_z\otimes{\bold I}_\tau$. The matrix ${\Psi}^\dagger_\bo$ contains the Bogoliubov operators $(\alpha_{\bo}^{\dg},~\beta_{\bo}^{\dg})$, $\mathcal{U}_\bo$ is a $2N\times 2N$ matrix ($N$ is the number of sublattice),  $\epsilon(\bo)=\text{diag}[\epsilon_\lambda(\bo),\epsilon_\lambda(-\bo)]$  and $\epsilon_\lambda(\bo)$ are the  eigenvalues. Using the fact that $\mathcal U_\bo^\dg=\eta \mathcal U_\bo^{-1}\eta$ and $\mathcal U_\bo\mathcal U_\bo^{-1}={\bf I}$, we have \bea \eta\mathcal{H}(\bo) \mathcal{U}_\bo= \mathcal{U}_\bo\eta\epsilon(\bo).\eea Thus, we need to diagonalize a non-Hermitian Bogoliubov Hamiltonian  $\mathcal{H}_B(\bo)=\eta\mathcal{H}(\bo)$, whose eigenvalues are $\eta\epsilon(\bo)$ and the corresponding eigenvectors are the columns of $\mathcal{U}_\bo$. The explicit form of $\mathcal{U}_\bo$ is given by
 \begin{align}
& \mathcal{U}_\bo= \begin{pmatrix}
  u_\bo& -v_\bo^* \\
-v_\bo&u_\bo^*\\  
 \end{pmatrix},
\end{align} 
where $u_\bo,~v_\bo$ are  $N\times N$ matrices that satisfy \bea|u_\bo|^2-|v_\bo|^2=1.\eea  The positive eigenvalues of
 \begin{align}
\mathcal{H}_B(\bo)&=[{ (v_A- m_A)\tau_+^0} +(v_B+m_B){\tau_-^0}] \otimes{\sigma}_z\nonumber\\& - v_{1}(\tau_+f_\bo +h.c.)\otimes{\sigma}_z- i v_{2}(\tau_+f_\bo +h.c.)\otimes\sigma_y.
\label{ham3}
\end{align}
are given by
\begin{align}
\epsilon_\lambda(\bo)=\epsilon_\lambda(-\bo)=\sqrt{\frac{(a_\bo+ \lambda b_\bo)}{2}},
\label{eig}
\end{align}
where
\begin{align}
a_\bo&=(m_A-v_A)^2+(m_B+v_B)^2+2( v_1^2- v_2^2)|f_\bo|^2,\\
b_\bo&=\sqrt{\alpha_+^2\alpha_-^2+4( v_{1}^2\alpha_-^2+ v_{2}^2\alpha_+^2)|f_\bo|^2},
\end{align}
with
$\alpha_-=(m_B+v_B)-(m_A-v_A)$ and $\alpha_+=(m_A-v_A)+(m_B+v_B)$.
At the Dirac points $\bold K_\pm$, the structure factor vanishes $f_\bo=0$, we find that the gap is given by
\begin{align}
\Delta_{gap} =\epsilon_+(\bold K_+)-\epsilon_-(\bold K_+)=|m_B-v_B|-|m_A+v_A|.
\end{align}
\section{Charge density wave insulator}
\label{appen2}
In this appendix, we study one of the phases of the model --- charge density wave insulator. 
The CDW insulator can be captured by setting $\mu=0$. Minimizing the classical energy we find $\theta_A=\arccos(\Delta/\Delta_c)$ and $\theta_B=\arccos(-\Delta/\Delta_c)$. A special case of CDW insulator is the fully polarized N\'eel state which occurs at $\Delta=\Delta_c$. Hence $\theta_A=0$ and $\theta_B=\pi$ with $m_A=\Delta_c$, $m_B=-\Delta_c$, $v_A=v_B=0$, and $v_1=0,~v_2=-2JS$. The Hamiltonian possesses an explicit analytical diagonalization. In the basis $\Psi_\bo^\dg=(c_{\bo A}^{\dg},~c_{-\bo B},~c_{\bo B}^\dg,~c_{-\bo A})$, Eq.~\ref{mat} takes the form
\begin{figure}[!]
\centering
\includegraphics[width=3.5in]{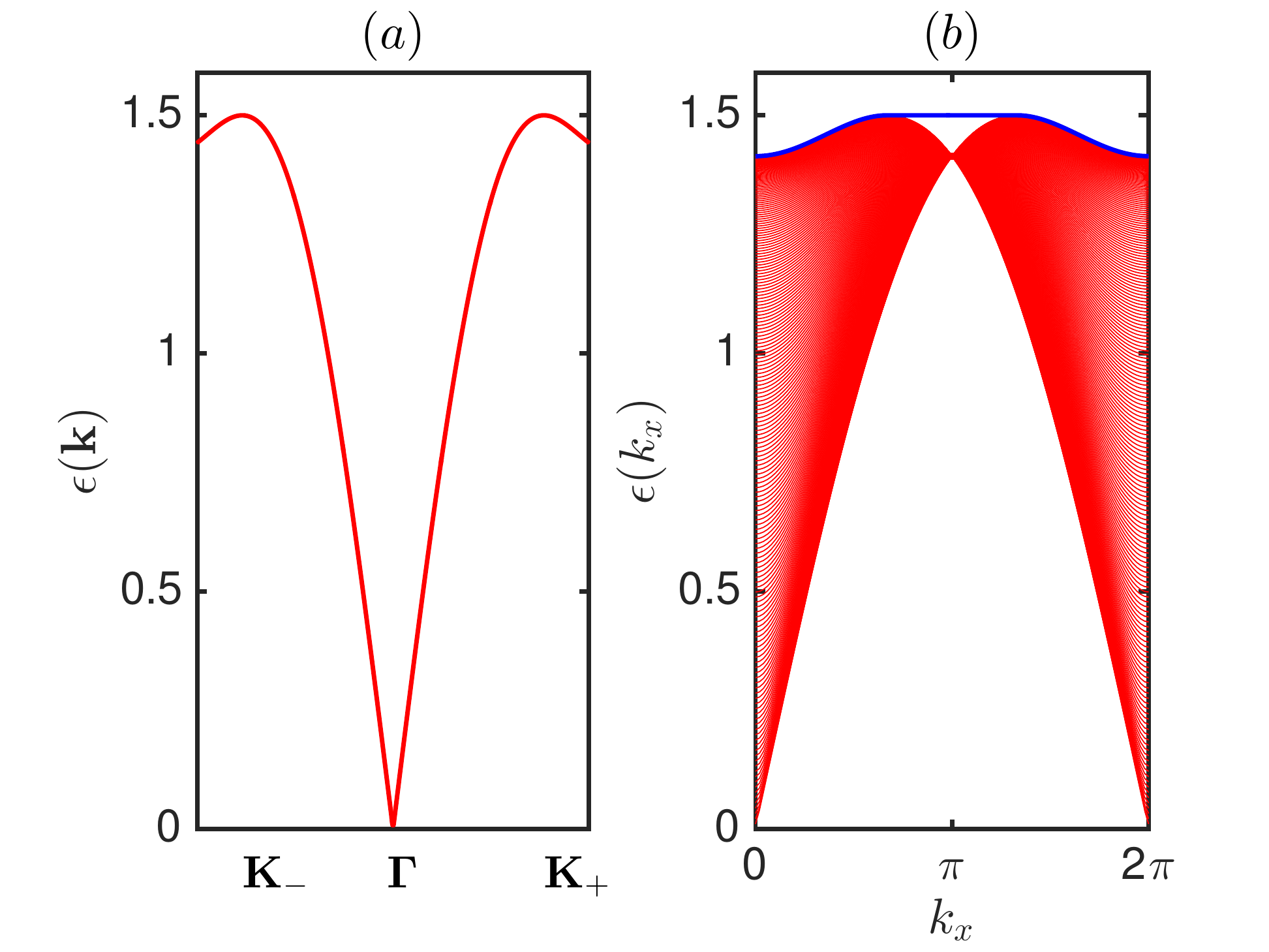}
\caption{Color online. Bosonic magnon band  $(a)$  and edge state (green solid line) $(b)$ at $\mu=0,~\Delta=\Delta_c$, corresponding to fully polarized N\'eel state.}
\label{Nband}
\end{figure}

\begin{figure}[!]
\centering
\includegraphics[width=3.5in]{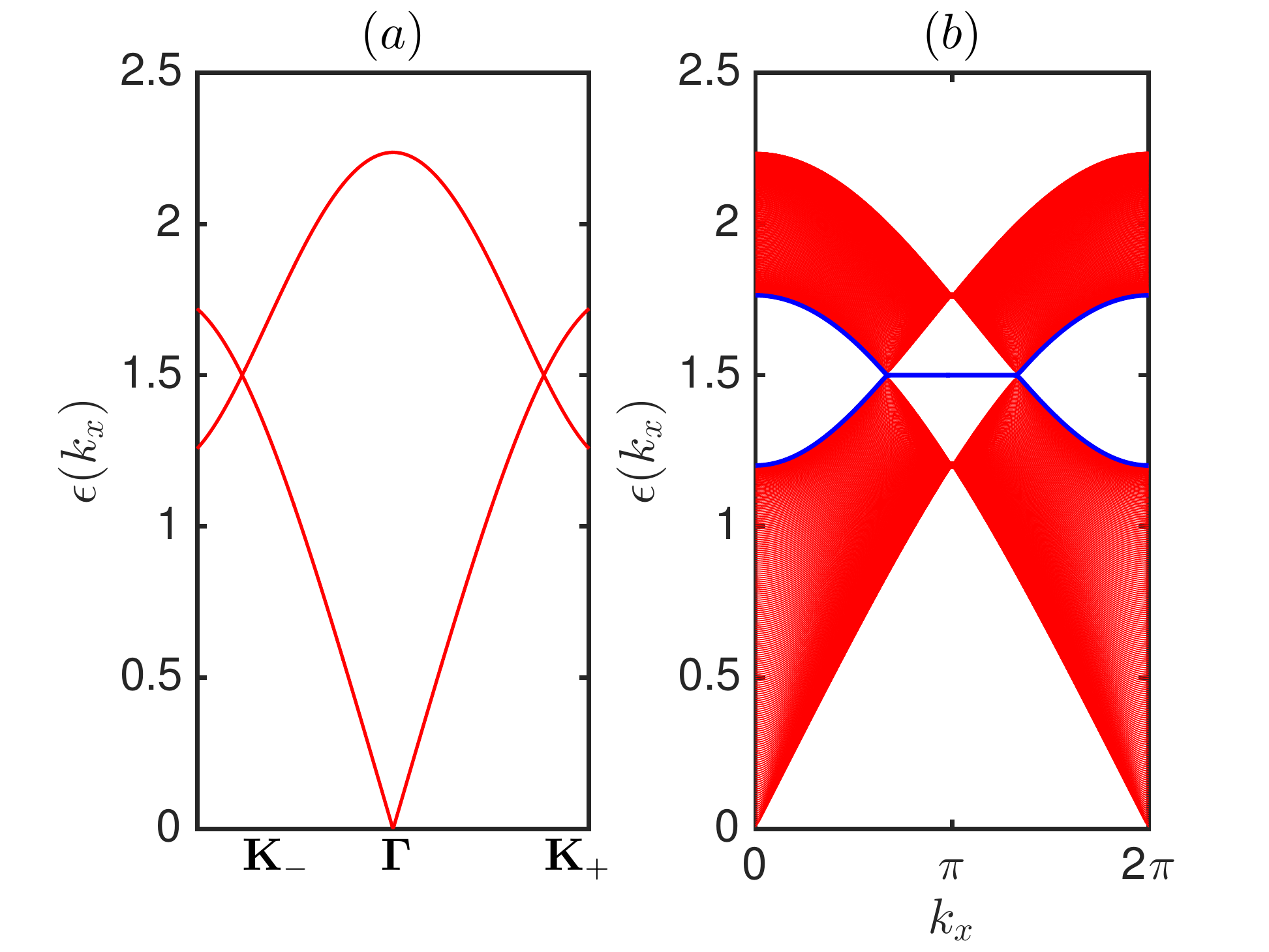}
\caption{Color online.  Bosonic magnon band  $(a)$  and edge state (green solid line) $(b)$ at $\mu=J,~\Delta=0$, corresponding to gapless SF phase.  }
\label{gapless_SF}
\end{figure}

\begin{align}
\mathcal{H}(\bo)=
\begin{pmatrix}
-\Delta_c&-v_2f_\bo& 0&0\\
-v_2f^*_\bo& -\Delta_c&0&0\\
0&0& -\Delta_c&-v_2f_\bo\\
0&0&-v_2f_\bo^*&-\Delta_c
\end{pmatrix}.
\label{mat1}
\end{align}
This is exactly the Heisenberg antiferromagnet on the honeycomb lattice. Since each block is doubly degenerate, we consider only block I, given by
\begin{align}
\mathcal{H}_I(\bo)=
\begin{pmatrix}
-\Delta_c&-v_2f_\bo&\\
-v_2f^*_\bo& -\Delta_c
\end{pmatrix}.
\label{mat2}
\end{align}
The matrix to be diagonalized is $\mathcal{H}_I^B(\bo)=\sigma_z\mathcal{H}_I(\bo)$. For magnon the eigenvalues are positive definite given by 
\begin{align}
\epsilon_I(\bo)=\sqrt{\Delta_c^2-v_2^2|f_\bo|^2}
\end{align}
The band structure is depicted in Fig.~\ref{Nband} (a). By solving the eigenvalue equation $\mathcal{H}_I^B {-v_\bo^*\choose u_\bo^*}=\epsilon_I(\bo){-v_\bo^*\choose u_\bo^*}$, we find
\begin{align}
u_\bo&=e^{i\phi_\bo}\cosh\lb\frac{\theta_\bo}{2}\rb,\quad
v_\bo=\sinh\lb\frac{\theta_\bo}{2}\rb,
\end{align}

where 
 \begin{align}
 \cosh\theta_\bo&=\frac{\Delta_c}{\epsilon_I(\bo)};~ \sinh\theta_\bo=\frac{|v_2f_\bo|}{\epsilon_I(\bo)};~
 \tan\phi({\bf k})= \frac{\textrm{Im} f_{\bold k}}{\textrm{Re} f_{\bold k}}.
 \end{align}
 
 The matrix $\mathcal U_\bo$  is given by 
 \begin{align}
 \mathcal U_\bo=\begin{pmatrix}
 e^{i\phi_\bo}\cosh\lb\frac{\theta_\bo}{2}\rb & -\sinh\lb\frac{\theta_\bo}{2}\rb\\
-\sinh\lb\frac{\theta_\bo}{2}\rb & e^{-i\phi_\bo}\cosh\lb\frac{\theta_\bo}{2}\rb
\end{pmatrix}.
 \end{align}
 The phase factor $\phi_\bo$ generates a nonzero Berry curvature defined in Eq.~\ref{chern2}, which can be reduced to a compact form
 \begin{align}
 \Omega_{\mu\nu}(\bo)=-2\text{Im}[\sigma_z\mathcal (\partial_{k_\mu}\mathcal U_\bo^\dg)\sigma_z(\partial_{k_\nu}\mathcal U_\bo)]
 \label{bc1}
 \end{align}
%where $\mu,\nu=x,y$ and $\mathcal A_\nu(\bo)=i\sigma_z\mathcal U_\bo^\dg\sigma_z(\partial_{k_\nu}\mathcal U_\bo),$ given by
%\begin{align}
%\mathcal A_\nu(\bo)=i\bigg[e^{-i\phi_\bo}\cosh\lb{\theta_\bo}/{2}\rb\partial_{k_\nu}[e^{i\phi_\bo}\cosh\lb{\theta_\bo}/{2}\rb]\nonumber\\-2\sinh\lb{\theta_\bo}/{2}\rb\partial_{k_\nu}[\sinh\lb{\theta_\bo}/{2}\rb]\label{bc2}\bigg ],
%\end{align}
%where $c.c$ is the complex conjugate of the phase term.
 The diagonal elements are given by
\begin{align}
 \Omega_{\mu\nu}^{11}(\bo)&=-2\text{Im}\bigg[\partial_{k_\mu}\lb e^{-i\phi_\bo}\cosh\lb\frac{\theta_\bo}{2}\rb\rb\nonumber\\&\times \partial_{k_\nu}\lb e^{i\phi_\bo}\cosh\lb\frac{\theta_\bo}{2}\rb\rb\bigg],\\\Omega_{\mu\nu}^{22}(\bo)&=-2\text{Im}\bigg[\partial_{k_\mu}\lb e^{i\phi_\bo}\cosh\lb\frac{\theta_\bo}{2}\rb\rb\nonumber\\&\times \partial_{k_\nu}\lb e^{-i\phi_\bo}\cosh\lb\frac{\theta_\bo}{2}\rb\rb\bigg].
\end{align}
These expressions can be reduced to
\begin{align}
 \Omega_{\mu\nu}^{11}(\bo)&=\frac{\sinh\theta_\bo }{2}[\partial_{k_\mu}\phi_\bo\partial_{k_\nu}\theta_\bo-\partial_{k_\nu}\phi_\bo\partial_{k_\mu}\theta_\bo ]\nonumber\\&=-\Omega_{\mu\nu}^{22}(\bo).
\end{align}
The partial derivatives can be simplified further, we obtain
\begin{align}
\partial_{k_\mu}\phi_\bo&= \frac{1}{|f_\bo|^2}[\text{Re}f_\bo\partial_{k_\mu}\text{Im}f_\bo-\text{Im}f_\bo\partial_{k_\mu}\text{Re}f_\bo],\\
\partial_{k_\mu}\theta_\bo&=\frac{|v_2|\Delta_c}{\epsilon^2(\bo)}\partial_{k_\mu}|f_\bo|.
\end{align}
Further simplification yields
\begin{align}
 \Omega_{xy}^{11}(\bo)=\frac{3\sqrt{3}\bigg[ \cos\lb\frac{\sqrt 3}{2}k_x\rb-\cos\lb\frac{3}{2}k_y\rb\bigg]\sin\lb\frac{\sqrt 3}{2}k_x\rb}{4\sqrt{2}\bigg[ 3-\cos\sqrt 3k_x-2\cos\lb\frac{\sqrt 3}{2}k_x\rb\cos\lb\frac{3}{2}k_y\rb\bigg]^{3/2}}.
 \end{align}
The Berry curvatures at the corner of the Brillouin  zone are given by
\begin{align}
 \Omega_{xy}^{11}({\bf K_\pm}) = \mp\frac{1}{8}.
 \end{align}
 This is consistent with the maxima and minima peaks of the positive Berry curvature shown above. It can be easily shown that the integration of the Berry curvature vanishes, hence the Chern number. However, there is an edge state propagating on the boundary of the system as shown in Fig.~\ref{Nband} (b).
\section{Superfluid phase}
In this appendix, we address the superfluid phase. 
As mentioned above, there are two  superfluid phases for $\mu <\Delta_c$, $\Delta=0$ (gapless) and $\mu <\Delta_c$, $\Delta<\Delta_c$ (gap $\sim\Delta$).  For the former, we have $\theta_A=\theta_B=\theta=\arccos(\mu/\Delta_c)$ and for the latter $\theta_A$ is not necessarily equal to $\theta_B$ leading to a gap $\sim\Delta$ at ${\bf K}_\pm$ as shown above in Fig.~\ref{band}(b). In the former case, $m_A=m_B=0$, $v_A=v_B=v_\theta=\Delta_c\sin^2\theta+\mu\cos\theta=\Delta_c$ and $v_{1,2}=JS(\cos^2\theta\pm 1)$. The eigenvalues Eq.~\ref{eig} yields
\begin{align}
\epsilon_\pm(\bo)&=\sqrt{[\Delta_c \pm v_1|f_\bo|]^2- (v_2f_\bo|)^2}\nonumber\\&
=2JS\sqrt{(3\pm |f_\bo|)(3\pm|f_\bo|\cos^2\theta )}.
\end{align}
At the Dirac points ${\bf K}_\pm$, we have $\epsilon_\pm(\bo)=|\Delta_c|=6JS$ and the system is gapless ($\Delta_{gap}=0$) as shown in Fig.~\ref{gapless_SF}(a). However, there exist an edge state connecting the two Dirac points as shown in Fig.~\ref{gapless_SF}(b). The spectrum disperses  linearly at two points in the Brillouin zone, ${\bf K}_\pm$ and ${\bf \Gamma}$. The former is the usually  Dirac points, i.e. linear touching of two bands $\epsilon_\pm(\bold q )= (6JS)[1\pm \frac{|\bold q|}{8}(3+\cos2\theta)] + \mathcal O(|\bold q|^2)$ ($\bold q=\bo-{\bf K}_\pm$), and the latter occur at the lower band $\epsilon_-(\bold q)= JS\sin\theta |\bold q| + \mathcal O(|\bold q|^2)$ ($\bold q=\bo-{\bf \Gamma}$). It corresponds to a Goldstone mode associated with breaking of U(1) symmetry.
\label{appen3}


\begin{thebibliography}{99}
\bibitem{yu6}
M. Z. Hasan and C. L. Kane, Rev. Mod. Phys. {\bf 82}, 3045 (2010).
\bibitem{yu7}
X.-L.  Qi and S.-C. Zhang, Rev. Mod. Phys. {\bf 83}, 1057 (2011).
\bibitem{yu1}
R. Yu {\it et al.}, Science, {\bf 329}, 61 (2010).
\bibitem{yu2}
 Y. L. Chen {\it et al.}, Science {\bf 329}, 659 (2010).
\bibitem{yu3}
 C.L. Kane and E.J. Mele, Phys. Rev. Lett. {\bf 95}, 146802 (2005); {\it ibid} Phys. Rev. Lett. {\bf 95}, 226801 (2005) .
\
\bibitem{yu4}
 J.E. Moore and L. Balents, Phys. Rev. B {\bf 75}, 121306 (2007).
 \bibitem{yu}
 L. Fu, C.L. Kane, and E.J. Mele, Phys. Rev. Lett. {\bf 98}, 106803 (2007).
 
\bibitem{yu5}
 D. Hsieh {\it et al.}, Nature {\bf 452}, 970 (2008).

\bibitem{yu8}
H.  Zhang {\it et al.}, Nature Phys. {\bf 5}, 438, (2009).
\bibitem{fdm}
F. D. M. Haldane, \prl {\bf 61}, 2015 (1988).
 \bibitem{sem}
 G. W. Semenoff, V. Semenoff, and F. Zhou, \prl {\bf 101}, 087204 (2008).
  \bibitem{sem1}
 W. Yao, S. A. Yang, and Q. Niu, \prl {\bf 102}, 096801 (2009).
  \bibitem{var1}
C. N. Varney {\it et al.,} \prb {\bf 82}, 115125  (2010).
 \bibitem{pa1}
I. Vasi\'c {\it et al.},  \prb {\bf 91}, 094502 (2015).
%\bibitem{jot}
%G. Jotzu {\it et al.}, Nature {\bf 515}, 237 (2014).
%\bibitem{pa}
%C.  Hickey, P.  Rath, A.  Paramekanti, \prb {\bf 91}, 134414 (2015).
%\bibitem{pa1}
%I. Vasi\'c {\it et al.},  \prb {\bf 91}, 094502 (2015).
%\bibitem{pa2}
%T. I. Vanhala {\it et al.}, arXiv:1512.08804.
%\bibitem{pa3}
%V. S. Arun, R. Sohal, C. Hickey, and A. Paramekanti, \prb {\bf 93}, 115110 (2016).
% \bibitem{iva}
%I. Vasi\'c, A. Petrescu, K. L.  Hur, and W. Hofstetter, \prb {\bf 91}, 094502 (2015).
 
     \bibitem{matq}
T. Matsubara and H. Matsuda,  Prog. Theor. Phys. {\bf 16}, 569 (1956).
\bibitem{alex0}
 H. Katsura, N. Nagaosa, and P. A. Lee,   \prl  {\bf 104},  066403 (2010).

 \bibitem{alex1}
 S.  Y. Onose  \textit {et al}.,  Science  { \bf 329}, 297 (2010).
  \bibitem{alex1a}
 Y. F. Wang {\it et al.,} \prl, {\bf 107}, 146803 (2011).
 \bibitem{alex2}
 R. Matsumoto and S. Murakami, \prl {\bf 106}, 197202 (2011); \prb {\bf 84}, 184406 (2011).
 \bibitem{alex4}
A.  Mook, J.  Henk, and I. Mertig, \prb {\bf 90}, 024412 (2014);  \prb {\bf 89}, 134409 (2014).
 \bibitem{alex5}
 H. Lee, J. H. Han, and P. A. Lee,   \prb  {\bf 91},  125413 (2015).
 \bibitem{alex5a}
 R. Chisnell \textit {et al}., \prl {\bf 115}, 147201  (2015).
 \bibitem{alex6} M.  Hirschberger \textit {et al}.,  \prl {\bf 115}, 106603 (2015).
\bibitem{sol}
S. A.  Owerre, J. Phys.: Condens. Matter 28, 386001 (2016).
 \bibitem{sol1}
S. A.  Owerre, J. Appl. Phys. {\bf 120}, 043903 (2016).
\bibitem{alex9}
 H. Guo \textit {et al}., \prb {\bf 93}, 121401(R) (2016).



\bibitem{ber}
  K. Bernardet {\it et al.}, {\prb  {\bf 65}, 104519 (2002)}.
  \bibitem{tom}
 T. Coletta, N. Laflorencie, and F. Mila, {\prb  {\bf 85}, 104421 (2012)}.
% \bibitem{mar}
%M.  Greiner {\it et al.}, Nature, {\bf 415}, 39 (2002).
%\bibitem{bec}
%C.  Becker {\it et al.}, New Journal of Physics, {\bf 12}, 065025 (2010).
%
 \bibitem{kle}
G.  Murthy, D.  Arovas, and A.  Auerbach, \prb {\bf 55}, 3104 (1997). 
 

 
 
  \bibitem{footnote} Also notice that this system cannot possesses magnon Hall effect \cite{alex1, alex0, alex2,alex5,alex4, sol1,sol, alex9, alex5a, alex6,  alex1a} because of the vanishing integrated Berry curvatures for each band.
   \bibitem{jot}
 {G. Jotzu} {\it et al},~ Nature {\bf 515}, {(2014)} {237}.

   \bibitem{jot2}
 Wei Zheng and Hui Zhai, Phys. Rev. A 89, 061603(R) (2014).
   \bibitem{jot1}
 Jiao Miao, Phys. Rev. A 92, 023632 (2015).
%  \bibitem{shin}
% R. Shindou et.al., Phys. Rev. B 87, 174427 (2013).
  % \bibitem{stro1} L. Sheng, D. N. Sheng, and C. S. Ting,  \prl {\bf 96}, 155901 (2006).
%  \bibitem{stro2} Y. Kagan, L. A. Maksimov, \prl  {\bf 100}, 145902 (2008).
%  \bibitem{alex7a}
%  L. Zhang {\it et al}., \prl {\bf 105}, 225901 (2010).
%\bibitem{stro3}  L. Zhang, J. Ren, J.-S. Wang, and B. Li, \prl {\bf 105}, 225901 (2010).
%\bibitem{stro4}  L. Zhang, J. Ren, J.-S. Wang, and B. Li, J. Phys. Condens. Matter
%{\bf 23}, 305402 (2011).
% \bibitem{alex2}
% R. Matsumoto and S. Murakami, \prl {\bf 106}, 197202 (2011); \prb {\bf 84}, 184406 (2011).
%% \bibitem{thou}
%%D. J. Thouless, M. Kohmoto, M. P. Nightingale, and M. den Nijs, Phys. Rev. Lett. {\bf 49}, 405  (1982); M. Kohmoto, Annals of Physics {\bf 160}, 343 (1985).
%
%\bibitem{alex4}
%A.  Mook, J.  Henk, and I. Mertig, \prb {\bf 90}, 024412 (2014);  \prb {\bf 89}, 134409 (2014).

%  \bibitem{alex7}
% A.  A. Kovalev and V.  Zyuzin, arXiv:1509.05847.
%
% \bibitem{jf}
% J. Fransson, A. M. Black-Schaffer, A.V. Balatsky,  { arXiv:1512.04902}.
%
%\bibitem{kkim}
%S. K. Kim \textit {et al}., { arXiv:1603.04827}.
%
%\bibitem{mar}
%M. P. A. Fisher {\it et al.}, \prb {\bf 40}, 546 (1989).
%\bibitem{mar1}
%D. Jaksch, {\it et al.}, \prl {\bf 81}, 3108 (1998).
%\bibitem{mar2}
%C. Orzel, {\it et al.}, Science {\bf 291}, 2386 (2001).
%\bibitem{mar3}
%M.  Greiner {\it et al.}, Nature, {\bf 415}, 39 (2002).
%\bibitem{mar4}
%L.-M. Duan, E. Demler, and M. D. Lukin, \prl {\bf 91}, 090402 (2003).
%\bibitem{mar5}
%C.  Becker {\it et al.}, New Journal of Physics, {\bf 12}, 065025 (2010).
%  
% \bibitem{alex8}
% Stefan Wessel,  \prl {\bf 75}, 174301 (2007).
% \bibitem{alex8a}
% J. Yu Gan,  \prl {\bf 75}, 214509 (2007).
%
%
%  \bibitem{ban}
% S.  Banerjee {\it et al.}, 	arXiv:1511.05282.
%
%  \bibitem{var}
%C. N. Varney {\it et al.}, {\prl {\bf 107}, 077201 (2011)}; C. N. Varney {\it et al.}; 	New J. Phys.  {\bf 14}, 115028 (2012); J. Carrasquilla {\it et al.} \prb {\bf 88}, 241109(R) (2013); Z.  Zhu, D. A. Huse, and S.  R. White, {\prl {\bf 111}, 257201 (2013)}.
%   \bibitem{alex10}
% D.P. Arovas and Assa Auerbach, \prb {\bf 38}, 316  (1988).
% \bibitem{alex11}
% A. A. Burkov and A. H. MacDonald, \prb {\bf 66}, 115320  (2002).


\end{thebibliography}
\end{document}